\documentclass[12pt,preprint]{aastex}

\usepackage{natbib}
\begin{document}

\title{GSC 07396--00759 = V4046 Sgr C[D]: a Wide-separation Companion to the Close T
  Tauri Binary System V4046 Sgr AB} 

\author{J. H.\ Kastner\altaffilmark{1}, G. G. Sacco\altaffilmark{1},
  R. Montez Jr.\altaffilmark{1}, D. P. Huenemoerder\altaffilmark{2},
  H. Shi\altaffilmark{1}, 
  E. Alecian\altaffilmark{3}, C. Argiroffi\altaffilmark{4,5}, M. Audard\altaffilmark{6,7},
  J. Bouvier\altaffilmark{8}, 
 F. Damiani\altaffilmark{5},
  J.-F. Donati\altaffilmark{9}, S. G. Gregory\altaffilmark{10},
  M. G\"{u}del\altaffilmark{11}, G. A. J. Hussain\altaffilmark{12}, A. Maggio\altaffilmark{5},
  T. Montmerle\altaffilmark{13} }

\altaffiltext{1}{Center for Imaging Science, Rochester Institute of
Technology, 54 Lomb Memorial Drive, Rochester NY 14623 USA
(jhk@cis.rit.edu)}
\altaffiltext{2}{MIT, Kavli Institute for Astrophysics and Space Research, 77 Massachusetts Avenue, Cambridge, MA 02139, USA}
\altaffiltext{3}{Observatoire de Paris, LESIA, 5, place Jules Janssen, F-92195 Meudon Principal Cedex, France}
\altaffiltext{4}{Dip. di Fisica, Univ. di Palermo, Piazza del
  Parlamento 1, 90134 Palermo, Italy}
\altaffiltext{5}{INAF - Osservatorio Astronomico di Palermo, Piazza del Parlamento 1, 90134 Palermo, Italy}
\altaffiltext{6}{ISDC Data Center for Astrophysics, University of
  Geneva, Ch. d'Ecogia 16, CH-1290 Versoix, Switzerland}
\altaffiltext{7}{Observatoire de Gen\'eve, University of Geneva,
  Ch. des Maillettes 51, 1290 Versoix, Switzerland}
\altaffiltext{8}{UJF-Grenoble 1 / CNRS-INSU, Institut de Plan\'etologie
  et d'Astrophysique de Grenoble (IPAG) UMR 5274, Grenoble, F-38041,
  France}
\altaffiltext{9}{IRAP-UMR 5277, CNRS \& Univ. de Toulouse, 14
  Av. E. Belin, F-31400 Toulouse, France}
\altaffiltext{10}{California Institute of Technology, MC 249-17,
  Pasadena, CA 91125 USA}
\altaffiltext{11}{University of Vienna, Department of Astronomy, T\"urkenschanzstrasse 17, 1180 Vienna, Austria}
\altaffiltext{12}{ESO, Karl-Schwarzschild-Strasse 2, 85748 Garching
  bei M\"{u}nchen, Germany}
\altaffiltext{13}{Institut d'Astrophysique de Paris, 98bis bd Arago, FR 75014 Paris, France}

\begin{abstract}
  We explore the possibility that GSC 07396--00759 (spectral type M1e)
  is a widely separated ($\sim$2.82$'$, or projected separation
  $\sim$12,350 AU) companion to the ``old'' (age $\sim$12 Myr)
  classical T Tauri binary system V4046 Sgr AB, as suggested by the
  proximity and similar space motions of the two systems. If the two
  systems are equidistant and coeval, then GSC 07396--00759, like V4046 Sgr AB, must
  be a spectroscopic binary with nearly equal-mass components, and
  V4046 Sgr must be at least
  $\sim$8 Myr old.  Analysis of a serendipitous Chandra X-ray gratings
  spectrum and light curve as well as XMM-Newton light curves and CCD
  spectra of GSC 07396--00759 obtained during long exposures targeting
  V4046 Sgr AB reveals a relatively hard ($T_X \sim 10^7$ K) X-ray
  spectrum, strong flaring, and relatively low-density plasma. These
  X-ray characteristics of GCS 07396--00759 are indicative of a high
  level of coronal activity, consistent with its apparent weak-lined T
  Tauri star status.  Interactions between V4046 Sgr AB and GCS
  07396--00759 when the two systems were more closely bound may be
  responsible for (a) their dissolution $\sim10^6$ yr ago, (b) the
  present tight, circular orbit of V4046 Sgr AB, and (c) the
  persistence of the gaseous circumbinary disk still orbiting V4046
  Sgr AB.
\end{abstract}
%\keywords{stars: circumstellar matter --- stars: early-type --- stars:
%  emission-line --- stars: supergiants --- Magellanic Clouds}

\section{Introduction}

Thanks to their proximity, individual members of the various nearby
($D\stackrel{<}{\sim}$100 pc), young (age $\sim$10--70 Myr) stellar
groups identified over the past $\sim$15 yr \citep[for a recent
review, see][]{2011ApJ...732...61Z} provide readily accessible
examples of, among other things, the late evolution of protoplanetary
disks \cite[e.g.,][and references therein]{2011ApJ...728...96A,2011ApJ...727...85H} and the early stages of evolution of exoplanetary
systems \citep[][]{2008Sci...322.1348M} and hierarchical binary star
systems \citep[][]{2008A&A...491..829K}.  In their tests of a method
to identify such comoving groups of young stars,
\citet{2006A&A...460..695T} established as a candidate member of the
$\beta$ Pic Moving Group ($\beta$PMG) the close binary classical T
Tauri system V4046 Sgr \citep[$P\sim2.4$
d;][and references
therein]{2004A&A...421.1159S}. Via the cluster traceback method,
\citet{2008hsf2.book..757T} estimated a distance of only 73 pc to
V4046 Sgr. Although the age of the $\beta$PMG is estimated to be
$\sim$12 Myr \citep[][and references therein]{2004ARA&A..42..685Z},
the twin members of this fascinating, short-period binary are both
evidently still accreting \citep{2004A&A...421.1159S} from a
relatively large and massive circumbinary disk of gas and dust
\citep{2008A&A...492..469K,2010ApJ...720.1684R}.

\citet{2006A&A...460..695T,2008hsf2.book..757T} also identified, as
another $\beta$PMG candidate, the star GSC 07396--00759
\citep[hereafter GSC0739, spectral type M1e or
M1.5e;][]{2006AJ....132..866R}, on the basis of its proximity to V4046
  Sgr (separation $2.82'$) and the similar radial
  velocities and similarly large photospheric Li abundances of the two systems 
\citep[the Li $\lambda$6708 equivalent
  widths of V4046 Sgr and GSC0739 are
440 and 200 m\AA, respectively;][]{2006AJ....132..866R,2009A&A...508..833D}.
% (Fig.~\ref{fig:Kimage}).
Given the low surface density of known $\beta$PMG members, it is
reasonable to hypothesize that GSC0739 is a distant (projected separation
$\sim$12.4 kAU) companion to the V4046 Sgr binary, as speculated by
\citet{2006A&A...460..695T}. In this respect the
relationship of GSC0739 to V4046 Sgr may be very much like that of 2M1235-39 to
the well-studied TW Hya Association (TWA) system HR 4796AB
\citep{2008A&A...491..829K}, whose primary is a young ($\beta$
Pic-like) A star with debris disk, or that of the brown dwarf
candidate TWA 28 to TW Hya itself \citep{2008A&A...489..825T}.

If GSC0739 and 2M1235-39 (= HR 4796C, the putative tertiary component
of HR 4796) are distant companions to the better-studied V4046 Sgr and
HR 4796, respectively, then their present orbital periods are
$\sim$10\% of the ($\sim$10 Myr) ages of these systems. Hence these
objects may offer unique insight into the orbital evolution
(and especially the dissolution) of young hierarchical binaries and,
perhaps, the likelihood of, and conditions necessary for, formation of
planets with circumbinary vs.\ circumstellar orbits.

Here, we briefly evaluate the available data concerning the space motions of
V4046 Sgr and GSC0730, and
we analyze optical/IR photometry and serendipitous X-ray
observations of GSC0739 (including new XMM-Newton 
X-ray data obtained during the course
of a coordinated observing campaign targeting V4046 Sgr; Argiroffi et
al.\ 2011). We use the results to
assess the likelihood that GSC0739 is indeed a distant companion to V4046
Sgr and to better ascertain the nature of the former. Finally, we consider
the implications if the two systems are (or were) physically bound.

\section{Data, Analysis, and Results}

\subsection{Proper motions and radial velocities}

The UCAC3 catalog \citep{2010AJ....139.2184Z} proper motions (PMs) listed for V4046 Sgr and
GSC0739  are ($\mu_\alpha, \mu_\delta) = (+3.3\pm1.7, -52.0\pm1.3$) mas yr$^{-1}$  and ($+1.1\pm2.8, -40.0\pm13.6$)
mas yr$^{-1}$, respectively, indicating that (given the UCAC3
  measurement errors) 
the PMs of the two
systems are indistinguishable\footnote{The identical PMs listed for the two systems by
\citet[][their Table 6]{2006A&A...460..695T} both
correspond to the Tycho catalog PM of V4046 Sgr AB, i.e., 
($+2.1\pm2.1, -54.5\pm2.3$) mas yr$^{-1}$; there is no Tycho catalog listing for GSC0739.}.
%The UCAC3 PM catalog data
%confirm that GSC0739 is the only star brighter than $J=12$ within
%$15'$ of V4046 Sgr that has similar PM components (i.e., RA and dec
%PMs within $\pm15$ mas yr$^{-1}$ of the PM components of V4046 Sgr).
%(Fig.~\ref{fig:Kimage}). 
The heliocentric radial
velocities ($V_{\rm helio}$) of the two systems are also very similar: \citet[][]{2006A&A...460..695T} 
list $V_{\rm helio} = -5.7$ km s$^{-1}$ for GSC0739, and CO
radio line measurements yield $V_{\rm helio} = -6.2\pm0.2$ km s$^{-1}$
for V4046 Sgr AB \citep[][]{2008A&A...492..469K,2010ApJ...720.1684R}\footnote{Recent results from
  high-resolution optical spectroscopy indicate that
the systemic velocity of V4046 Sgr AB may vary by $\sim0.5$ km
s$^{-1}$ (Donati et al.\ 2011).}. 
%The
%latter is similar to the value of $-6.9$ km s$^{-1}$ listed for V4046
%Sgr in \citet[][]{2006A&A...460..695T}.

\subsection{Optical/IR SED and comparison with pre-MS tracks}

By matching Kurucz model atmospheres \citep{1993yCat.6039....0K}
appropriate for late-type main sequence
stars (i.e., $\log{g} = 4.5$) 
%see http://www.stsci.edu/hst/observatory/cdbs/k93models.html
to optical/near-IR photometry available in SIMBAD\footnote{The SIMBAD
  database is maintained and operated at CDS, Strasbourg, France.}
for GSC0739 (Table~\ref{tbl:phot}), we deduce a photospheric
  effective temperature 3500$\pm$100 K, consistent with the M1--1.5
  spectral type previously determined for GSC0739
  \citep{2006A&A...460..695T,2006AJ....132..866R}. From the Kurucz
  model normalization, and assuming the distance to GSC0739 is 73 pc
  (i.e., that GSC0739 snd V4046 Sgr are equidistant), we determine
  a bolometric luminosity $L_{\rm bol}= 0.15$ $L_\odot$ (with an
  approximate uncertainty, dominated by the uncertainty in distance, 
of $\sim25$\%). This estimate is consistent with that
  obtained from the J magnitude and V--K color of
  GSC0739, $L_{\rm bol}= 0.17$ $L_\odot$, following the method
  described in \citet{1995ApJS..101..117K}.  
%The former result is consistent, within the
%uncertainties, with the V4046 Sgr AB system $L_{\rm bol}$ determined
%by \citet{2000IAUS..200P..28Q}, after accounting for the slightly
%larger distance to V4046 Sgr assumed by those authors.

We used the foregoing results for GSC0739 and adopted the luminosity
and temperature results reported for the individual components of
V4046 Sgr AB by Donati et al.\ (2011) to place
these systems on theoretical pre-MS tracks
(Fig.~\ref{fig:preMStracks}). The comparison illustrates that the two
systems can be both equidistant and coeval only if GSC0739, like V4046 Sgr AB, consists of
two components with nearly equal luminosities. Indeed, GSC0739 is
flagged as a possible spectroscopic binary by
\citet{2006A&A...460..695T}. The positions of the GSC0739 and V4046
Sgr AB systems with respect to the overlaid model evolutionary tracks
indicate ages of between $\sim$10 and $\sim$20 Myr, if GSC0739 is a
binary. This is consistent with independent (kinematic and pre-MS
evolutionary track) age estimates of $\sim$12 Myr for the $\beta$PMG
\citep[see review in][]{2004ARA&A..42..685Z}. On the other hand, if
GSC0739 is single, then Fig.~\ref{fig:preMStracks} indicates that its
age is in the range $\sim$3--10 Myr, i.e., somewhat young for a
$\beta$PMG member.
%supporting the conclusion \citep{2006A&A...460..695T,2008hsf2.book..757T} that both V4046 Sgr AB and GSC0739 are members of this
%association.

\subsection{X-ray light curves and spectra}

\begin{figure}[htb]
  \centering
\includegraphics[width=5in,angle=0]{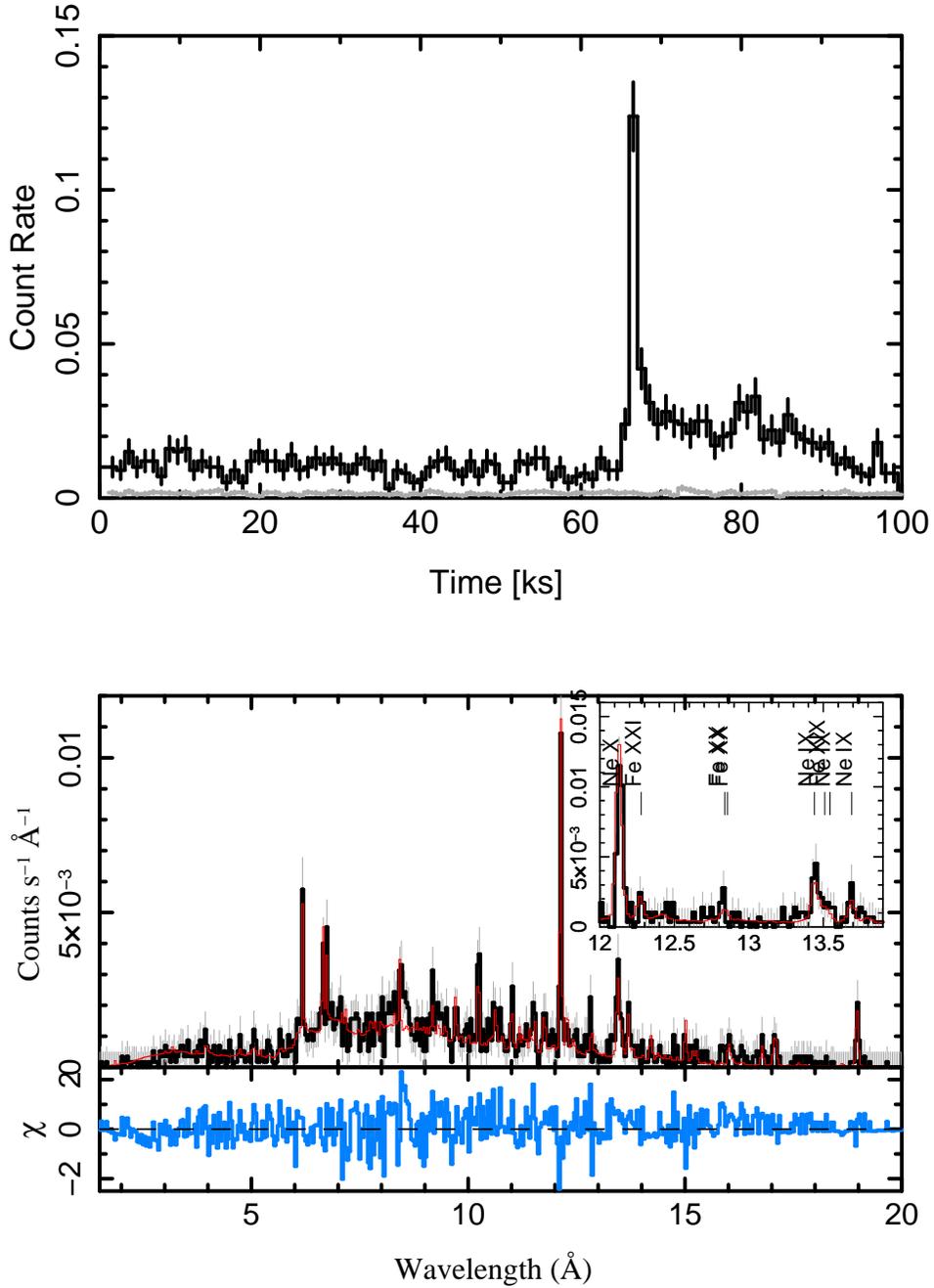}
%\vspace{-.5in}
\caption{{\it Top:} Chandra/HETGS light curve of GSC0739, covering the longer ($\sim$100 ks) continuous
  observing segment of the two-segment 145 ks exposure. Black: source;
  grey: background. {\it Bottom:} Chandra/HETGS spectrum of GSC0739
  (black, with grey error bars) overlaid with
  best-fit APED plasma model (red) in which the emission measure
  distribution is assumed to follow a power law (see text). The lower
  panel shows the residuals of the fit. The inset is a blowup of the 12.0--13.9 \AA\ region, 
  illustrating the strong emission lines of Ne {\sc x} and Ne {\sc ix} as well
  as weak lines of highly ionized Fe.}
\label{fig:Chandra} 
\end{figure}

An off-axis Chandra/HETGS gratings spectrum of GSC0739 (total exposure
144.6 ks; OBSIDs 5422, 6265; PI: Herczeg) was serendipitously obtained
in 2005 Aug.\ during an observation targeting V4046 Sgr
\citep{2006A&A...459L..29G}. The GSC0739 X-ray spectral data are
uncontaminated by V4046 Sgr AB, but the spectral resolution and the exposure 
of the GSC0739 source are somewhat
compromised by its ($\sim$$3'$) off-axis position. We extracted and
combined the medium-energy gratings (MEG) and high-energy gratings
(HEG) spectra for the full $\sim$145 ks exposure duration
and generated a light
curve covering the first (longer) of the two exposure segments; the resulting
light curve and spectrum are displayed in
Fig.~\ref{fig:Chandra}. The total (HEG+MEG) count rate,
averaged over the entire 144.6 ks exposure duration, was
$0.014\,\mathrm{counts\, s^{-1}}$.

\begin{figure}[htb]
  \centering
\includegraphics[width=6in,angle=0]{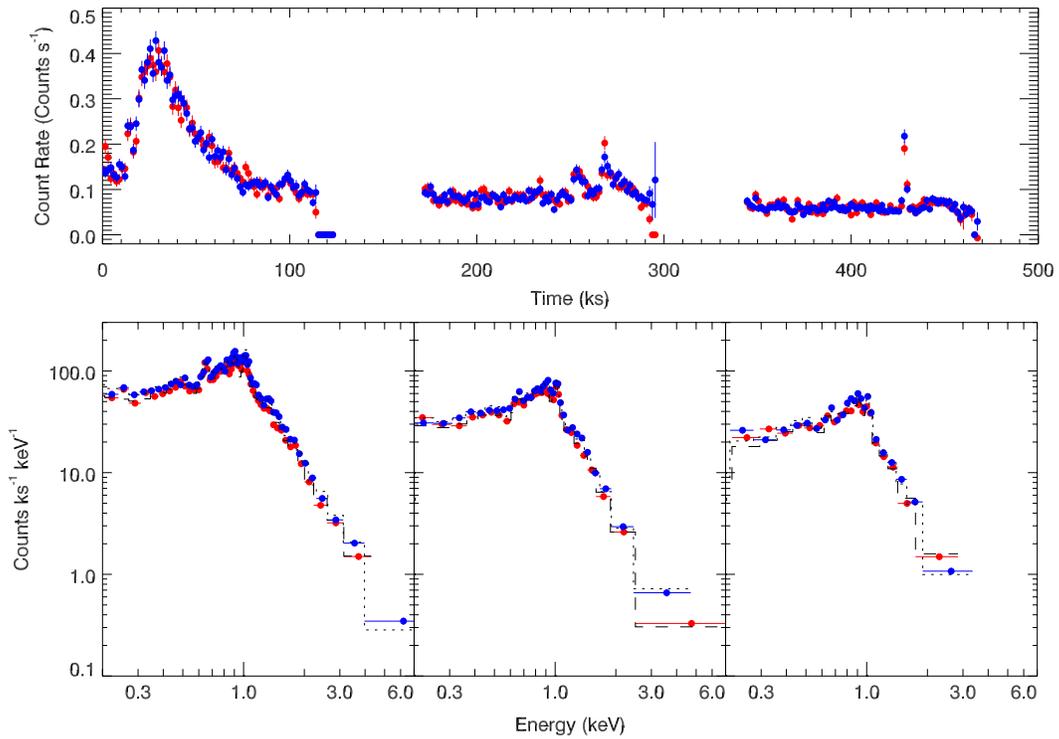}
\vspace{-1.5in}
\caption{{\it Top:} XMM/EPIC MOS light curve of GSC0739. Binsize is
  1500 s. {\it Bottom:} MOS spectra corresponding to the three
  $\sim$100 ks observing intervals illustrated in the light curve in
  the top panel, with best-fit two-component thermal plasma models
  overlaid. In all panels, red points indicate MOS1 data and blue
  points indicate MOS 2 data.}
\label{fig:XMM} 
\end{figure}

XMM European Photon Imaging Camera (EPIC) 
data were serendipitously obtained for GSC0739 in 2009 Sept.,
again during an observation (3 exposures of $\sim$100 ks) targeting
V4046 Sgr (Argiroffi et al.\ 2011). As the source
fell along a bad column of EPIC's pn detector, we only analyzed data
from the two MOS detectors (the source also fell near the edge of the
active area of MOS1, but these data were only slightly compromised).
We used the XMM Scientific Analysis System
(SAS\footnote{http://xmm.esa.int/sas} version 10.0.0) to extract MOS1
and MOS2 light curves for the full ($\sim$5 day) duration of the
$3\times100$ ks exposure (top panel of Fig.~\ref{fig:XMM}) as well as
MOS1 and MOS2 CCD spectra and responses for each individual $\sim$100
ks exposure interval (bottom panels of Fig.~\ref{fig:XMM}). Calibrations were performed using the
current calibration files (CCF) from release note 271, 21-Dec-2010. Combined
MOS1+MOS2 count rates are listed in Table~\ref{tbl:fitParamsTF}.

The Chandra and XMM light curves reveal strong X-ray flaring at
GSC0739. An impulsive flare with amplitude $\sim$10 times the
quiescent flux level and exponential decay is observed starting at
about 65 ks into the first $\sim$100 ks segment of the Chandra
observation (Fig.~\ref{fig:Chandra}, top panel). A flare of similar shape (with
amplitude $\sim$4 times quiescent) is observed near the beginning of
the overall $\sim$5-day XMM exposure, and its decay appears to last
the entire duration of the first  $\sim$100 ks exposure segment; a smaller
flare (and several small count rate spikes) are also seen in the second and third
XMM exposure segments (Fig.~\ref{fig:XMM}, top).

Table~\ref{tbl:fitParamsTF} and Fig.~\ref{fig:XMM} (bottom panels)
summarize the results of fits of variable-abundance (see below)
absorbed two-component thermal plasma models
%to the full 145 ks exposure Chandra/HETGS spectrum and
to spectra extracted from the three $\sim$100 ks XMM/EPIC (MOS)
exposure intervals. A similar model fit to Chandra/HETGS data (not
shown) yields similar results --- i.e., a prominent hard component
with $T_X \sim$ 20 MK, and a weaker soft component with $T_X \sim$
5--7 MK --- despite the different instrumentation; we conclude that,
overall, the plasma physical conditions did not change markedly
between 2005 and 2009. Nevertheless, the strong short-term variability
of the source is apparent in the XMM fit results (Table~\ref{tbl:fitParamsTF}), which indicate a
steady decline in the emission measure of the hard component and,
possibly, a slight decrease in $T_X$ as the observation progressed
and the strong flare faded (the source may have suffered from photon
pileup during the first exposure interval).

Individual line intensities in the Chandra/HETGS data are not
particularly well fit by the foregoing simple (two-component) model,
as expected given the continuum of plasma temperatures that is likely
present in the X-ray-emitting region. Hence, we performed global
fitting of a broken power law emission measure distribution model
(using isis\footnote{http://space.mit.edu/cxc/isis/} and ATOMDB
v2.0\footnote{http://www.atomdb.org/}), such as used by \citet{RTV} or
\citet{Peres:01} for modeling coronal loops, to the gratings data.
The model is specified by an emission measure normalization, the peak
temperature ($T_{peak}$), the power law slopes below ($\alpha$) and
above ($\beta$) the peak temperature, and the elemental abundances.
The interstellar absorption was assumed to be negligible (a safe
assumption given the very modest absorption determined from the fits
to XMM data; Table~\ref{tbl:fitParamsTF}).  To account for the
instrumental off-axis PSF, we used a broadening equivalent to a
Gaussian turbulent velocity of $600\,\mathrm{km\,s^{-1}}$. The
resulting best fit model is overlaid on the Chandra/HETGS spectum in
Fig.~\ref{fig:Chandra} (bottom panel); the best-fit model parameters
are $T_{peak} = 11.2\,\mathrm{MK}$ ($0.97\,\mathrm{keV})$, $\alpha =
0.2$, $\beta = -1.8$, and an emission measure of
$3.3\times10^{52}\,\mathrm{cm^{-3}}$.  The best-fit abundance ratios
(relative to solar) are Ne/O $=1.8$, Mg/O $=0.27$, Si/O $=0.6$, and
Fe/O $=0.27$ (the two-component plasma model fits to the XMM/MOS data
also indicate an elevated Ne/O ratio and depressed Fe/O ratio). These
Ne/O and Fe/O ratios are typical of coronally active stars \citep[see
review in][]{2010SSRv..157...37T}. Analysis of X-ray gratings spectra
of V4046 Sgr AB likewise indicates enhanced Ne/O and low Fe/O,
although the Ne/O enhancement is more extreme \citep[a factor $\sim$5
larger than solar;][]{2006A&A...459L..29G} in
the case of this (actively accreting) system.

%The resonance/intercombination/forbidden ({\it rif}) line triplet of
%He-like Ne {\sc ix} at $\sim13.5$ \AA, which is diagnostic of plasma
%density, is potentially indicative of the presence or absence of
%ongoing accretion in T Tauri stars. Specifically, the {\it rif} line
%ratios within He-like species as measured for actively accreting
%(classical) T Tauri stars often reveal high plasma densities
%($\sim10^{11-12}$ cm$^{-3}$, as
%estimated from their rather small {\it f/i} ratios) that are larger
%than those generally associated with stellar coronae, indicative of an
%X-ray origin in accretion shocks
%\citet[e.g.,][]{kastner02,huenemoerder08,sacco10}; whereas the {\it
%  rif} ratios measured for nonaccreting (weak-lined) T Tauri stars are
%characteristic of lower densities ($\stackrel{<}{\sim}10^{10}$ cm$^{-3}$, based
%on {\it f/i} ratios), consistent with a coronal origin for the X-rays
%\citet{kastner04}. 
Although the resolution of the Chandra/HETGS spectrum of GSC0739
suffers from degradation due to the off-axis position of the source,
it is apparent that the forbidden to intercombination line ratio
within the He-like Ne {\sc ix} line complex (at $\sim$13.5 \AA) is
large (Fig.~\ref{fig:Chandra}, inset  in bottom panel), indicative of lower-density
($\stackrel{<}{\sim}$10$^{10}$ cm$^{-3}$)
plasma. Fig.~\ref{fig:Chandra} (bottom panel) also illustrates the relatively
large Ne {\sc x} to Ne {\sc ix} line ratios of GSC0739, reflecting the
dominance of relatively hot ($\stackrel{>}{\sim}$10 MK) plasma in its
emission measure distribution (Table~\ref{tbl:fitParamsTF}). These
density- and temperature-sensitive Ne line ratios are
consistent with coronal X-ray emission and, hence, a weak-lined T
Tauri classification for this source \citep[see, e.g.,
][]{2004ApJ...605L..49K,2007ApJ...671..592H}. Both Ne line
spectral diagnostics also stand in stark
contrast to those of V4046 Sgr AB; the latter displays Ne (and O) line
ratios indicative of a significant emission contribution from cooler
($\sim$3 MK) plasma at much higher density \citep[$\sim$10$^{12}$
cm$^{-3}$;][Argiroffi et al.\ 2011]{2006A&A...459L..29G}, as expected
if much of the X-ray emission (essentially, all of the soft component)
is generated in accretion shocks \citep[][and references
therein]{2010A&A...522A..55S}.

\section{Discussion and Conclusions}

The available proper motion and radial velocity data (\S 2.1) support
the hypothesis that V4046 Sgr AB and GSC0739 are comoving systems.
Hence -- provided GSC0739 is also a close binary (\S 3.2) -- the two
systems likely constitute a hierarchical multiple and, in the
following, we refer to GSC0739 as V4046 Sgr C[D]. 

\subsection{The nature of V4046 Sgr C[D]}

Based on its lack of H$\alpha$ emission \citep[H$\alpha$ equivalent
width of +3.1 \AA;][]{2006AJ....132..866R}, 
V4046 Sgr C[D] appears to be best classified as a
weak-lined T Tauri system (wTTS); i.e., unlike V4046 Sgr AB, %\citep{2004A&A...421.1159S}, 
the system shows no evidence for ongoing accretion.
The prominent hard ($T_X \sim 10^7$ K) X-ray spectral component,
strong X-ray flaring, and relatively low-density X-ray-emitting plasma of V4046
Sgr C[D] (\S 2.3) --- all of which are indicative of a high level of coronal
activity --- are fully consistent with such a wTTS classification.

\subsection{V4046 Sgr AB and C[D]: constraints on system age and origin}

If the V4046 Sgr AB and C[D] systems are indeed bound, they are only
weakly so, given their very large projected separation of $\sim$12,350
AU (assuming $D=73$ pc). Assuming the systems are still bound, the
orbital period of V4046 Sgr C[D] with respect to V4046 Sgr AB would be
of order $\sim10^6$ yr, i.e., a significant fraction of their
estimated ages. This system would therefore appear to be the most
loosely bound hierarchical binary system known in the
$\beta$PMG. However, two similarly loosely bound binaries have been
identified thus far in the TWA: TWA 11AB and TWA 11C, projected
separation $\sim$13500 AU \citep{2008A&A...491..829K} and TW Hya (TWA
1) and TWA 28, projected separation $\sim$41000 AU
\citep{2008A&A...489..825T}. The separations of these nearby,
  young hierarchical multiple systems are comparable to those of the
  most loosely bound binaries in the solar neighborhood \citep[the
  separation distribution of nearby field binaries appears to be
  truncated at $\sim$20,000 AU;][]{1990AJ....100.1968C}. The (very
low-mass) TWA 30A and 30B star/disk systems, which are separated on
the sky by $\sim$3400 AU \citep{2010AJ....140.1486L}, are also very
weakly bound.

\begin{figure}[htb]
  \centering
\includegraphics[width=6.5in,angle=0]{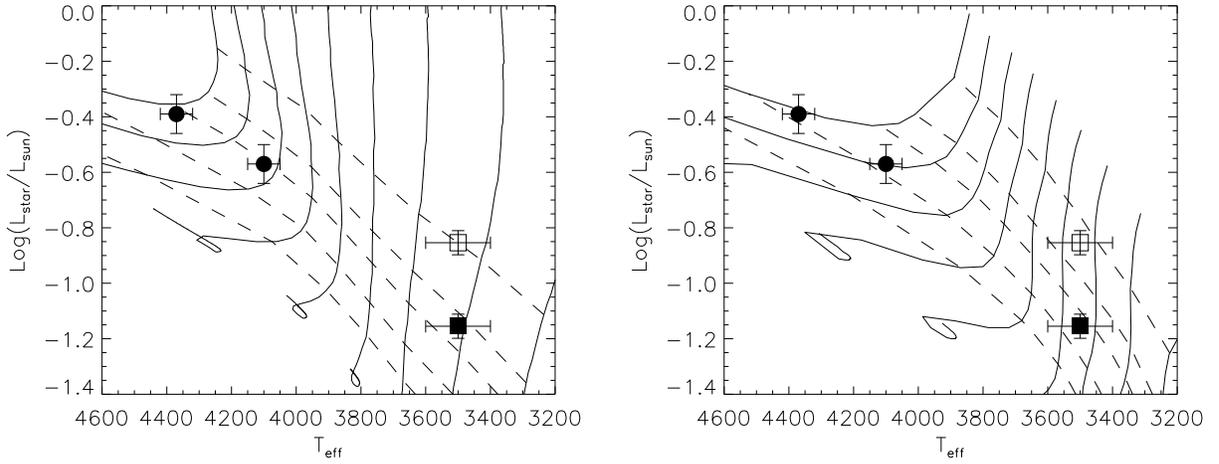}
\caption{The HR diagram positions of V4046 Sgr AB (circles; luminosity and temperature data from
  Donati et al.\ 2011) and GSC0739 (squares; see \S 2.2)
  overlaid on pre-MS tracks from \citet[][left panel]{2000A&A...358..593S} and
  \citet[][right panel]{1998A&A...337..403B}.  
  The empty square indicates the observed position of GSC0739; the
  filled square indicates its position assuming it is a binary system composed of a pair of identical stars.
In each panel, the evolutionary tracks (solid lines) correspond to
masses ranging from 0.2 to 1.0 $M_\odot$ (in intervals of 0.1 $M_\odot$,
from right to left), while the isochrones (dashed lines) correspond
to ages of 5, 10, 15, 25 and 40 Myr
(from top to bottom).}
\label{fig:preMStracks} 
\end{figure}

However, the
analysis in \S 2.2 strongly implies that V4046 Sgr AB and the putative
V4046 Sgr C[D] could only be coeval if the latter, like the former, is
a nearly-equal-components binary --- consistent with previous
suspicions concerning this system \citep{2006A&A...460..695T}.
Furthermore, if the two systems indeed formed together, then the
position of V4046 Sgr C[D] in Fig.~\ref{fig:preMStracks} appears to
place a conservative lower limit of $\sim$8 Myr on the age of the V4046 Sgr
multiple system, while --- on the basis of this comparison of V4046 Sgr C[D] with pre-MS tracks
alone, and given the present uncertainties in the stellar parameters ---
the system may be as old as $\sim$25 Myr.

%It is also worth noting (see below) that the apparent divergence of
%V4046 Sgr AB and C[D] indicated (respectively) by their Tycho and
%UCAC2 proper motions indicates a trace-back time of only $\sim$10000
%yr for convergence of the two systems. However, this timescale is much
%smaller than the orbital period of V4046 Sgr C[D] with respect to
%V4046 Sgr AB assuming the systems are still bound (i.e., $\sim10^6$
%yr).  Furthermore, the PMs do not produce a convergence point that
%coincides in space and time. Hence, there is reason to doubt this
%simple trace-back time calculation.

%It is worth considering whether the presence of the wide-separation companion(s)
%V4046 Sgr C[D] may be related to --- indeed may be responsible for ---
%the close, circular orbit of the V4046 Sgr AB binary \citep[$P\sim$2.4
%d, $e<0.01$;][]{2004A&A...421.1159S}. 

Numerical simulations of hierarchical triple star systems demonstrate
that the action of Kozai oscillations, in combination with tidal
friction, shrinks and circularizes the orbit of the inner binary
\citep{2007ApJ...669.1298F}.  Although \citet{2007ApJ...669.1298F}
find that the fraction of circularized, close (few-day period)
binaries with tertiaries as distant as V4046 Sgr C[D] should be
negligibly small, they also note that $\sim$40\% of hierarchical
triple systems are ``lost'' from their numerical sample, due to
dissolution. Perhaps V4046 Sgr AB and C[D] constitute such a (recently
dissolved) hierarchical multiple system that, at a slightly
  earlier epoch, resembled closer, young hierarchical binaries (such
  as the $\sim$8 Myr-old TWA systems HD 98800 and Hen 3-600). If so,
  the dissolution of V4046 Sgr AB and C[D] evidently occurred
$\sim$10$^6$ yr ago and left behind a close binary with a nearly
circular orbit \citep[V4046 Sgr AB has an eccentricity
$e<0.01$;][]{2004A&A...421.1159S}.  

Given that V4046 Sgr has apparently well outlived the ``canonical''
few-Myr timescale for Jovian planet formation \citep[e.g.,][and
references therein]{2009ApJ...698....1C}, one might further speculate
that the presence of V4046 Sgr C[D] may explain the longevity of
  the circumbinary disk around V4046 Sgr AB --- just as the
  hierarchical natures of the young multiple systems HD 98800 and Hen
  3-600 may explain the persistence and observed properties of the
  circumbinary dust disk in each of these systems \citep[see,
  e.g.,][and references
  therein]{2001ApJ...549..590P,2007ApJ...664.1176F,2010ApJ...710..462A}. Although
  it remains to work out the details in all of these examples, in the
  specific case of V4046 Sgr it is possible that dynamical
  interactions with component(s) C[D] may have inhibited the
  formation of gas giant and ice giant planets in circumbinary orbits
  around components AB, thereby preserving the gaseous, circumbinary
disk from which this close binary system is still accreting.

\acknowledgments{\it This research was supported by NASA Goddard Space
  Flight Center XMM-Newton Guest Observer Facility and NASA
  Astrophysics Data Analysis Program grants to RIT (NASA grant numbers
  NNX09AT15G and NNX09AC96G, respectively). DPH was supported by NASA
  through the Smithsonian Astrophysical Observatory (SAO) contract
  SV3-73016 for the Chandra X-Ray Center and Science Instruments. MA
  acknowledges support from the Swiss National Science Foundation
  (grants PP002--110504 and PP00P2--130188). The authors thank
  B. Zuckerman for useful discussions, the anonymous referee for a
  thorough and insightful review, and Emily
  Thompson (West Irondequoit [NY] High School) for contributions
  to the catalog research reported herein.}

%\bibliography{refs}{}

\begin{table}[h!]
\footnotesize
\caption{\sc GSC0739: Photometry$^a$}
\label{tbl:phot}
\begin{center}
\begin{tabular}{ccccccc}
\hline
\hline
%Object &
B & V & R & I & J & H & K \\
\hline
%V4046 Sgr$^b$ & 11.6: &   10.8: &   10.2:  &   9.5:  &     8.07 (0.02)  &
%7.44 (0.05) &   7.25 (0.02) \\
%GSC0739 & 
14.14 &  12.78   & 12.13  &  10.64  &  9.44 (0.02)  &  8.77 (0.04)
& 8.54 (0.02) \\
\hline
\end{tabular}
\end{center}
a) Photometric data compiled from SIMBAD 
\citep[sources: Tycho and UCAC3 catalogs, and][]{2006A&A...460..695T}. 
Photometric uncertainties available only for 2MASS data.
%b) V4046 Sgr varies by a few tenths of a magnitude in the
%visible region \citep[e.g.,][]{2000IAUS..200P..28Q}. \\
\end{table}

\begin{table}[ht]
\footnotesize
\caption{\sc GSC0739: XMM X-ray spectral fit results$^a$}
\label{tbl:fitParamsTF} 
\begin{center}
\begin{tabular}{lccccccccc}
\hline
\hline
(obs.\ & $C^b$ & $N_H$ & $T_1$ & EM$_1^c$ & $T_2$ & EM$_2^c$ & $L_x$
& $\log{L_x/L_{\rm bol}}$\\
    interval)           & (s$^{-1}$)   &  ($10^{19}$ cm$^{-2}$)  & (MK)   &
               ($10^{52}$ cm$^{-3}$) & (MK) &  ($10^{52}$ cm$^{-3}$) &  (erg cm$^{-2}$ s$^{-1}$) & \\
\hline
%CXO/HETGS & {\bf [???]} & {\bf [???]}  & 7.5 & 0.9 & 22 &
%1.5 & 9.0$\times10^{-13}$  & $-2.99$ \\
(1) & 0.24 & 0.0--3.0 & 5.2--5.6 & 3.1--3.3 & 21--23 &
3.3--3.5 & $9.6\times10^{29}$  & $-2.76$ \\
(2) & 0.12 & 2.0--8.0 & 5.3--5.8 & 2.4--2.6 & 17--20 &
1.1--1.3 & $3.8\times10^{29}$   & $-3.17$ \\
%(3) & 0.081 & 2.5--2.9 & 5.2--5.6 & 2.8--3.1 & 14--27 &
%0.38--0.6 & 2.9 & $-3.28$ \\
(3) & 0.081 & 3.0$^d$ & 5.2--5.6 & 2.8--3.1 & 14--27 &
0.38--0.6 & $2.9\times10^{29}$  & $-3.28$ \\
\hline
\end{tabular}
\end{center}

NOTES: a) Fit to absorbed two-component, variable-abundance (APED) plasma models, with parameter ranges corresponding to 90\% confidence intervals.
%b) Count rates for Chandra/HETGS are combined MEG+HEG 1st order; for XMM/EPIC, combined %MOS1+MOS2.
b) Combined MOS1+MOS2 count rate.
c) Plasma component emission measure.
d) Parameter fixed during fitting.
\normalsize
\end{table}

\end{document}